# Electrochemical Thermodynamics, Kinetics, and Hysteresis in Energy Materials: Focusing on the Solid Side

Keyvan Malaie

Institute of Biochemistry, University of Greifswald, Felix-Hausdorff-Str. 4, 17489 Greifswald, Germany

E-mail: keyvan.malaie@uni-greifswald.de

## Abstract

Bulk electrochemical phase transitions (EPTs) are the cornerstone of most modern electrochemical technologies, underlying many energy storage and electrocatalytic systems. Nonetheless, the fundamental mechanisms governing EPTs in solid-to-solid systems remain only partially understood because they involve complex interactions between phase transitions and electrochemical reactions. This mini-review introduces the thermodynamics of EPTs based on the general framework of phase transitions and mixtures, followed by a discussion of electrochemical hysteresis and kinetics in EPTs. Finally, using recent insights into the EPTS in $Ni(OH)_2$, $LiFePO_4$, and $MnO_2$ materials, the importance of Cyclic Voltammetry (CV) modelling in discerning underlying reaction mechanisms is highlighted. This mini-review inspires fundamental research into solid-to-solid EPTs for improving the performance of current energy storage materials.

## Introduction

EPTs underpin numerous technologies such as electrodeposition, batteries, pseudocapacitors, electrochromic devices, and electrolyzers [1–4]. They can be limited to one or a few monolayers of atoms on the electrode surface, such as underpotential deposition of hydrogen/metals or surface restructuring [5–7] or occur within the electrode bulk, such as electrochemical ion (de)insertion, electrodeposition, and conversion reactions [8–10].

EPTs share some features of well-established phase transitions and mixtures, e.g., water vaporization or water-oil mixtures. This similarity arises because ETPs can be viewed as mixing ions with crystal sites on the electrode or mixing oxidant (O) and reductant (R) species inside a solid medium. However, EPTs differ from phase mixtures and phase transitions in certain aspects, because they are “reactive mixtures”, i.e., phase transition and reaction at the same time, which is coupled with electron and ion transfer across the electrode interface [11,12].

It is commonly believed that the mechanism of electrochemical phase transition follows the nucleation and growth model [13,14]. Hence, according to the classical and atomic theory of phase formation, in analogy to the latent heat of vaporization, the formation of a new phase requires nucleation work [15]. The nucleation work is required to create new surfaces until the surface and the bulk forces balance each other. A full discussion of the mechanism of electrochemical phase formation can be found in the references [13,16–18].

A commonly accepted view about the kinetics of EPTs is that at the beginning of the reaction, there are plenty of lattice sites for adsorption/insertion. Thus, the rate of EPT should be high at the initial stages. In the later stages, the reaction rate slows down due to the depletion of the available lattice sites. A simple model based on this statistical view is the Kolmogorov–Johnson–Mehl–Avrami (KJMA) theorem. It makes several simplifications and assumes homogeneous stochastic nucleation over a surface and a constant (or predefined) growth rate [19]. Accordingly, the fraction of the transformed phase ($f$) versus time ($t$) can be expressed phenomenologically as $f=1-\exp(-kt^n)$. However, its two parameters, $k$ and $n$, are not easy to interpret in modeling EPTs. In this model, $n$ determines the shape of the transition or growth mechanism, and $k$ is interpreted as a constant related to the rate of phase transition.

Scharifker and Motsaney (S-M) have used the Avrami equation with $n = 1$ to model the potentiostatic current transients of electrodeposition under diffusion limitation. According to the S-M model, if the initial number of active crystal sites is large/small, the nucleation is instantaneous/progressive, respectively [13]. In the later stages of electrodeposition, the current becomes limited by diffusion resulting in a peak in the galvanostatic current transient. The S-M model has been used successfully for the electrodeposition of mercury ions on platinum and is used in electrodeposition studies of other metals and even solid-to-solid systems [13]. However, the model parameters, $k$, and the initial number of active sites can depend on overpotential and other factors, leading to difficulties in their physical interpretation and deviation from experiments in electroplating studies [18,20,21]. In addition, whether EPT processes proceed via complete stochastic nucleation on surfaces according to Avrami and S-M models is open to question. Non-stochastic nucleation could be, for instance, due to imperfect surfaces, the presence of initial nuclei, or lattice mismatch. If nucleation phenomena were completely random, they should always start at different locations on the electrode surface by repeating electrochemical cycles, which needs to be verified. In 1997, Scholz et al. demonstrated through thermodynamic modeling of CV of solid-solid systems that the observed CV features of EPTs, such as peak separation and shape, may be governed by the miscibility gaps rather than nucleation work [22–24].

Herein, we first describe the thermodynamics of phase transition, i.e., vaporization. Next, we obtain the electrochemical signals of EPTs from the corresponding free energy curves in order to track down the electrochemical features of EPTs via their GCD and CV curves. Finally, we describe the unconventional electrochemical kinetics of EPTs in light of recent studies on $Ni(OH)_2$, $LiFePO_4$, and $MnO_2$.

## Results and Discussion

## Phase Transition

Phase transitions can be generally described based on the free energy per mole ($G_m$) at different temperatures ($T$) and pressures ($P$) described as $\mathrm{d}G_\mathrm{m} = -S_\mathrm{m}\,\mathrm{d}T + V_\mathrm{m}\mathrm{d}P$, where $S_m$ and $V_m$ are the molar entropy and volume of the phase. Fig. 1a displays the free energies of two different phases, for instance, a liquid and its vapor, at different $T$ at constant $P$. The curve with the steeper slope represents the vapor phase due to its larger entropy. Fig. 1b displays the entropy change associated with the transition from liquid to the vapor phase. This type of transition, which is accompanied by a discontinuity in entropy, is called a first-order transition, which originates from the latent heat. Most electrochemical phase transitions are classified as first-order phase transitions. Alternatively, if the free energy curves in Fig. 1a are rather similar, the entropy change of the phase transition would be continuous as displayed in Fig. 1c, and it is called a second-order transition. An example of a second-order transition is the transition from ferromagnetic to paramagnetic behavior.

Fig. 1d exhibits the equilibrium $T$-$P$ curve, also called the coexistence curve, for the liquid-vapor system. The behavior of phase transition near the critical point can be described by Landau's theory. After the critical point, the system is homogeneous and its free energy as a function of $x$ (an order parameter such as volume or density) is plotted in Fig. 1e. For a point on the coexistence curve, the system is heterogeneous, i.e., consists of two phases, and its free energy diagram is shown in Fig. 1f. $x_{trans}$ signifies the volumes at which the transition between a one-phase to a two-phase system takes place.

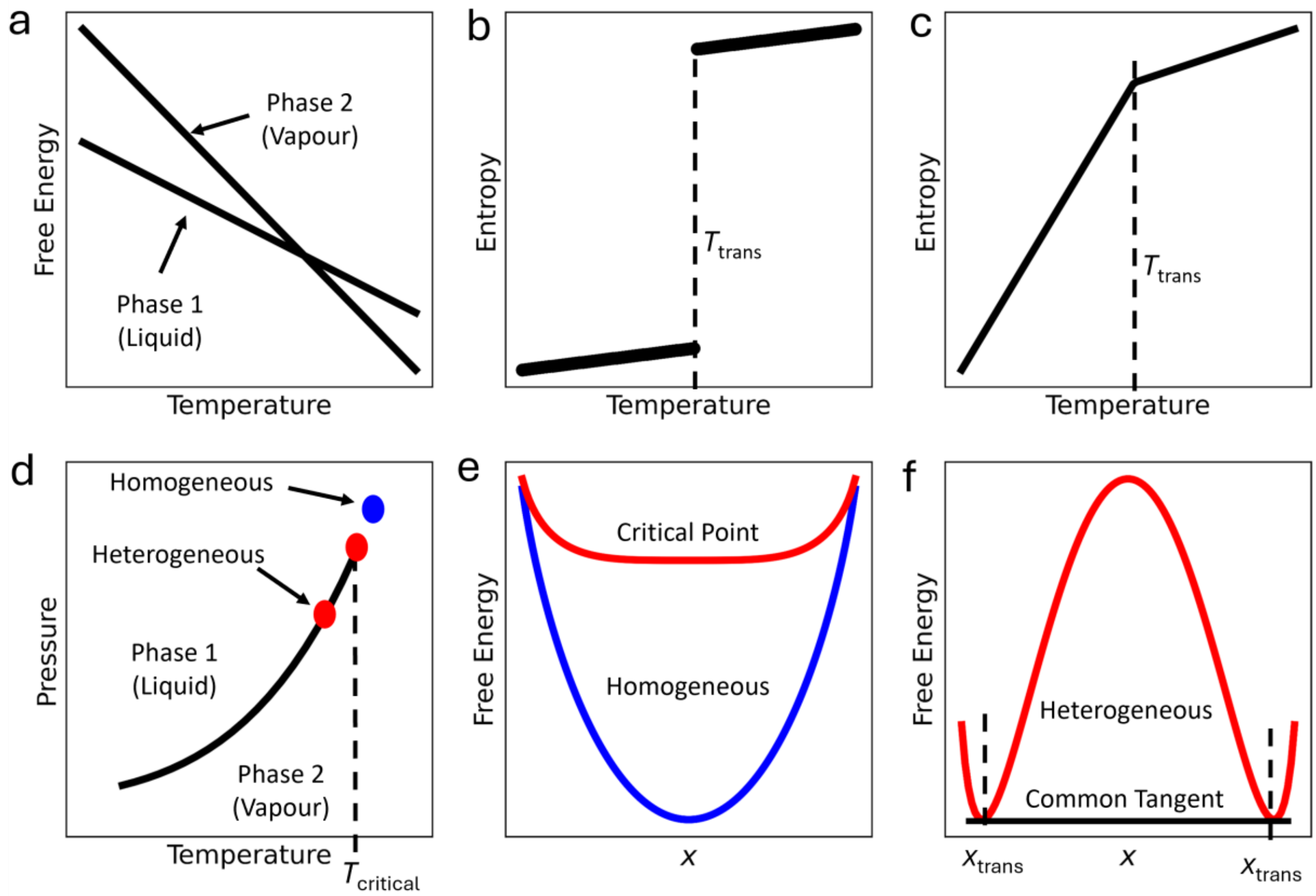

**Fig. 1 a)** Free energy change of two phases at constant pressure, **b)** A first-order transition **c)** A second-order transition **d)** Equilibrium phase diagram of a typical liquid/vapor system based on the Clausius-Clapeyron equation, and **e)** and **f)** show the free energy profiles corresponding to the points in Fig. 1d.

These free energy plots (Fig. 1 e and f) also apply to inert phase mixtures such as the water-oil mixture, where *x* denotes the composition or mole fraction. For this two-component system, the free energy plots can be obtained via mixing entropy and enthalpy [25]. The region between the two minima in Fig. 1f is also known as the miscibility gap. If a homogeneous phase initially exists in this region it separates into two phases to minimize its free energy. This double-minima energy pathway is usually not accessible in electrochemical systems because they can easily become unstable under current/voltage fluctuations, causing them to go through the common tangent. Thus, this thermodynamic instability makes the free energy curves below the critical point similar to the one at the critical point.

## Diffusion-Free EPT

In EPTs, phase transition is coupled with ion and electron transfer at the electrode interfaces. The coupling means that transferring one mole of electrons and one mole of ions across the interface(s) is energetically the same process as converting one mole of a redox phase into the other. Therefore, in principle, the free energy diagrams (Fig. 1e and f) can be used to derive and describe the electrochemical signals of EPTs, as well. For this purpose, the configurational entropy corresponding to the Free energy – *x* curves in Fig. 1 e and f is computed. Next, the equilibrium potential, *E*, can be obtained according to $\frac{\partial G_m}{\partial x} = \mu = -nFE$, where the symbols denote chemical potential, electron number, and Faraday constant, respectively. In this way, the following Nernst equation can be obtained for the homogeneous system (eq. 1) [25]:

$$E = E_{\mathrm{f}} + \frac{RT}{n_e F} \ln \frac{x}{1-x} \qquad (1)$$

$$E_{\mathrm{f}} = E^{0'} - \frac{2.302RT}{n_e F} \mathrm{pH} \qquad (2)$$

Where $E_f$ is a pH-dependent formal potential defined by eq. 2, corresponding to *x* = 0.5. Fig. 2a, the blue and red curves, exhibit diagrams of equilibrium potential versus *x* for a homogeneous and heterogeneous EPT obtained from the corresponding free energy curves in Fig. 1e. These curves are commonly encountered in galvanostatic charge-discharge tests of battery materials without and with phase separation, respectively. Fig. 2b displays a plot of formal potential for different solution pH, also called the Pourbaix diagram, for a typical metal oxide. The Pourbaix plots can give a very good first estimate of the type of EPT that operates under certain *E* and pH ranges. For instance, via obtaining experimental *E*-pH plots for the ε-$MnO_2$ under (quasi)reversible conditions, it was verified that the dominant operating EPT in alkaline solutions is proton (de)insertion and in acidic solutions both the proton-(de)insertion and the $MnO_2/Mn^{2+}$ reactions are verified [26].

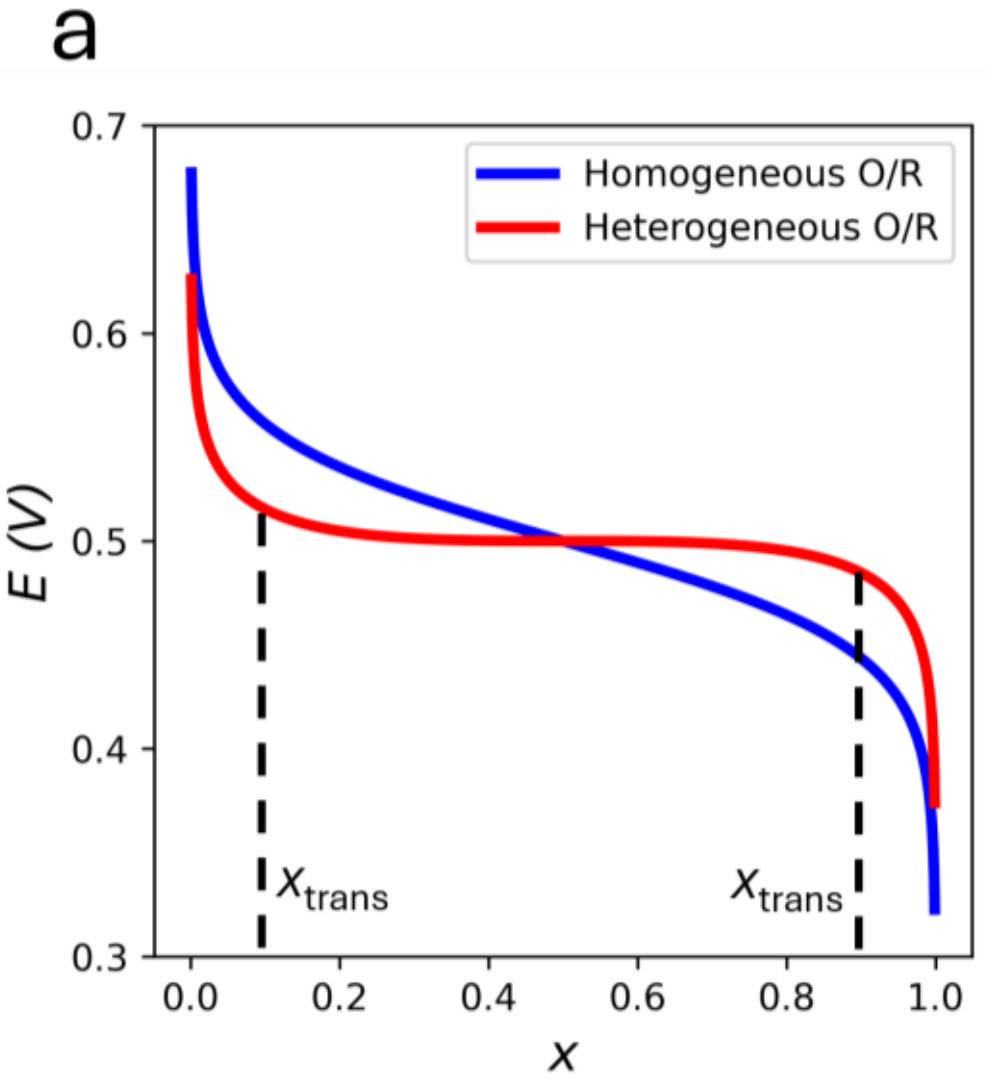

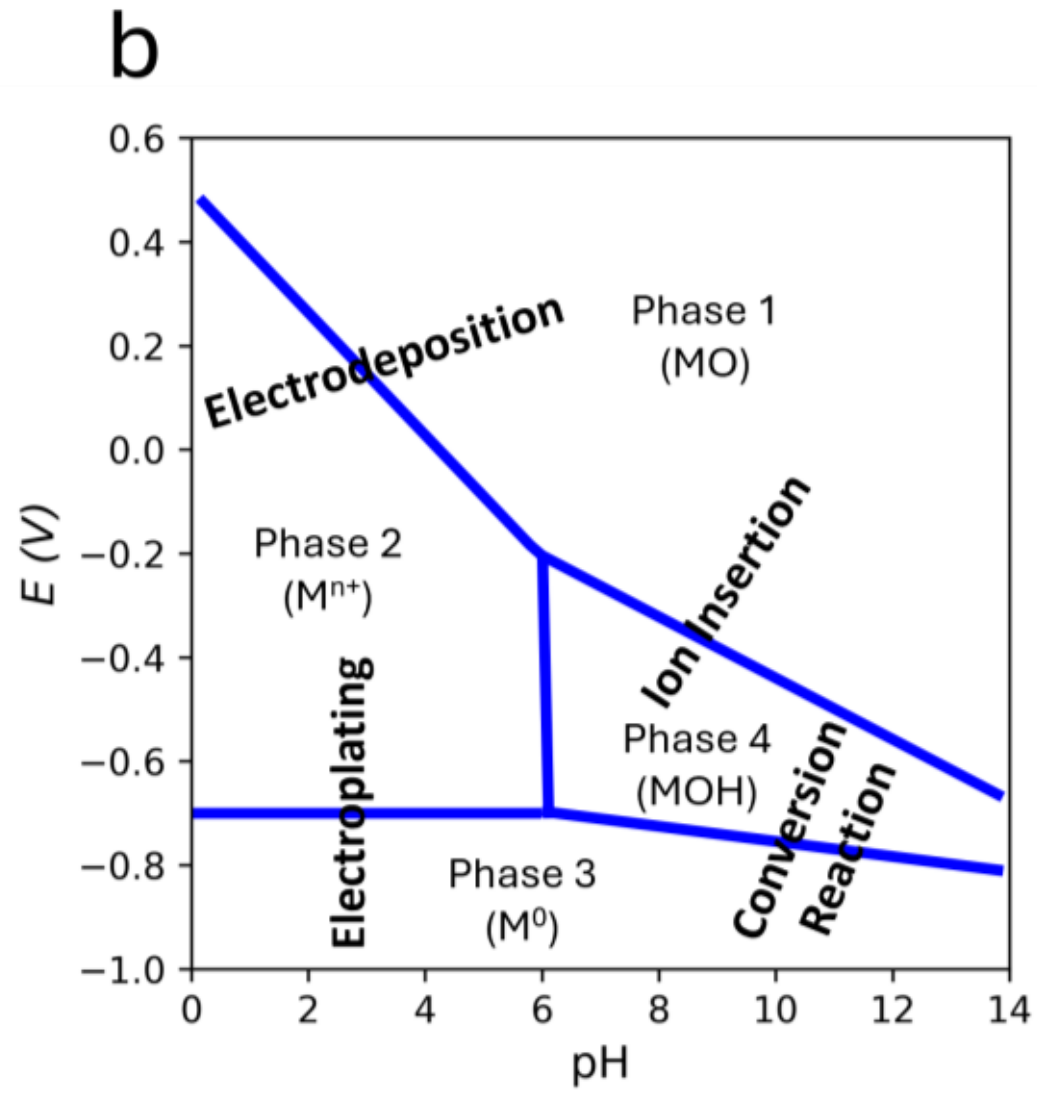

**Fig. 2 a)** Potential profiles of homogenous and heterogeneous O/R systems. **b)** A simplified phase diagram often encountered for metal oxides and the operating EPTs. (Water oxidation and reduction are not shown)

It is possible to use the as-obtained *E-x* plots (Fig. 2a) to derive thermodynamic CV curves for EPTs. In Fig. 3a, the blue and red CV curves are obtained from the corresponding *E-x* plots for EPTs under reaction control (no diffusion limitation). The narrow CV peaks correspond to a heterogeneous O/R phase transition. The broad CV curves with a full width at half maximum of about 90 mV correspond to an ideal homogenous EPT.

Often, EPTs are more complicated than those in Fig. 3a mainly due to phase separation during reaction. As a result, EPTs can easily show electrochemical hysteresis similar to magnetic hysteresis. Fig. 3b exhibits the representative CV cycle of a diffusion-free EPTs with strong (red curve) and weak hysteresis (blue curve). The oxidation and reduction potentials appear at different locations for this EPT. This hysteresis is accounted for in the free energy diagrams by including a term for the surface energy between the O and R phases. Accordingly, the surface energy for oxidation and reduction are different because oxidation and reduction go through different reaction pathways (reaction asymmetry), resulting in different energy landscapes [25]. In other words, when EPTs occur close to or below the critical point (Fig. 1d), the relative contribution of solid solution and phase separation pathways depends on the direction of the reaction. A reason for this complication is that the EPT cannot pass through the common tangent to the other side and phase boundaries remain at the end of oxidation/reduction time. It is noted here, that this hysteresis effect is not shared by the phase diagrams displayed in Fig. 1, because water always vaporizes and condenses at 100 °C at 1 atm, regardless of direction.

A comparison of these hysteretic CV curves (Fig. 3b) with the experimental CV curves of the $CoOOH/CoO_2$ and $Ni(OH)_2/NiOOH$ indicates that the phase separation and hysteresis originate from the different ionic radii of the metallic centers of O and R [25]. A similar observation can also be made for the widely studied phase separation in the $LiFePO_4$ cathode [27], in which the ionic radius of Fe can change by about 0.14 Å [28] during charge and discharge.

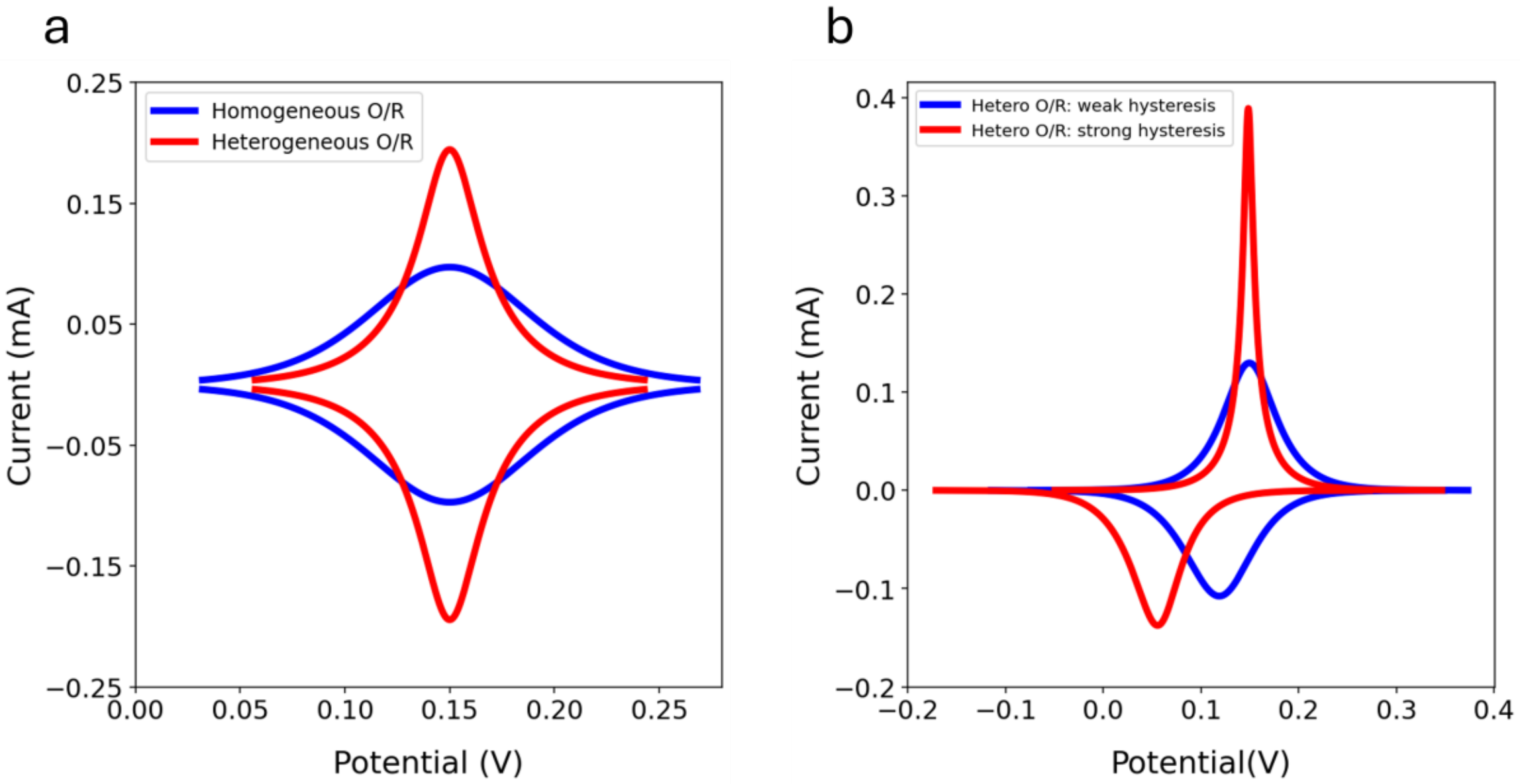

**Fig. 3** Thermodynamic CVs of EPT **a)** reaction-controlled EPTs, **b)** reaction-controlled homogeneous and heterogeneous EPTs with different levels of electrochemical hysteresis

We have recently introduced a nonthermodynamic model based on autocatalytic growth to describe the kinetics and mechanism of EPTs [29]. In this model, the rate of the reaction is not determined by stochastic nucleation and diffusion-limited growth, as suggested by the S-M model, but it is determined by the interactions between the O and R species. The reaction rate depends on the different stages ($x$) of particle growth, making it a size-dependent growth rate equal to $kx(1-x)^m$ as displayed in Fig. 4a. If the interaction between O and R crystal units are 1 to 1, i.e., $m$=1, the EPT is homogeneous as represented with the grey curve. In addition, for homogeneous EPTs, the interaction parameter, $k$, is identical for both oxidation and reduction directions. When $m$>1, the EPT is heterogeneous and involves phase separation (red curve). An inherent outcome of phase separation in this model is that the EPT is autocatalytic in the direction of phase formation (ion deinsertion) and autoinhibitory in the direction of phase dissolution (ion insertion).

This simple model is used to explain the (de)protonation mechanism of $Ni(OH)_2$ thin films during CV[30]. Fig. 4b (red) exhibits the first two CV cycles of $Ni(OH)_2$ thin film in 0.1 KOH at a scan rate of 2 mV/s and its comparison with the heterogeneous model. The sharp rise of the oxidation peak (an asymmetric peak) and its higher current compared to the reduction peak is due to the strong autocatalytic deprotonation of $Ni(OH)_2$ to $NiOOH/NiO_2$. The fact that the second oxidation peak appears at lower potentials is due to the residual $NiOOH/NiO_2$ remaining from the first cycle. Thus, the formation of $NiOOH/NiO_2$ particles promotes the deprotonation rate, as expected for an autocatalytic reaction. This effect of the initial amount of the product on the reaction rate is similar to the effect of nucleation overpotential on reaction rate in the electrodeposition of metals [31].

Fig. 4c (red curves) shows the second CV cycles of the same $Ni(OH0)_2$ at two different scan rates at 2 and 5 mV/s. Interestingly, the effect of scan rate for the deprotonation and protonation peaks are different. At a scan rate of 2 mV/s, the autocatalytic model (blue curves) simulates the CV curve with $k_{deinsert}/k_{insert}$ = 2.5, but at a scan rate of 5 mV/s this parameter needs to be reduced to 1.4 to simulate the CV curve. The lower value of this parameter indicates less phase separation during deprotonation at higher scan rates, and hence, lower reaction asymmetry. Bazant et al. [32] have for the first time reported this effect of charge-discharge rate on the reaction pathway via the galvanostatic curves for $LiFePO_4$ (de)lithiation [33].

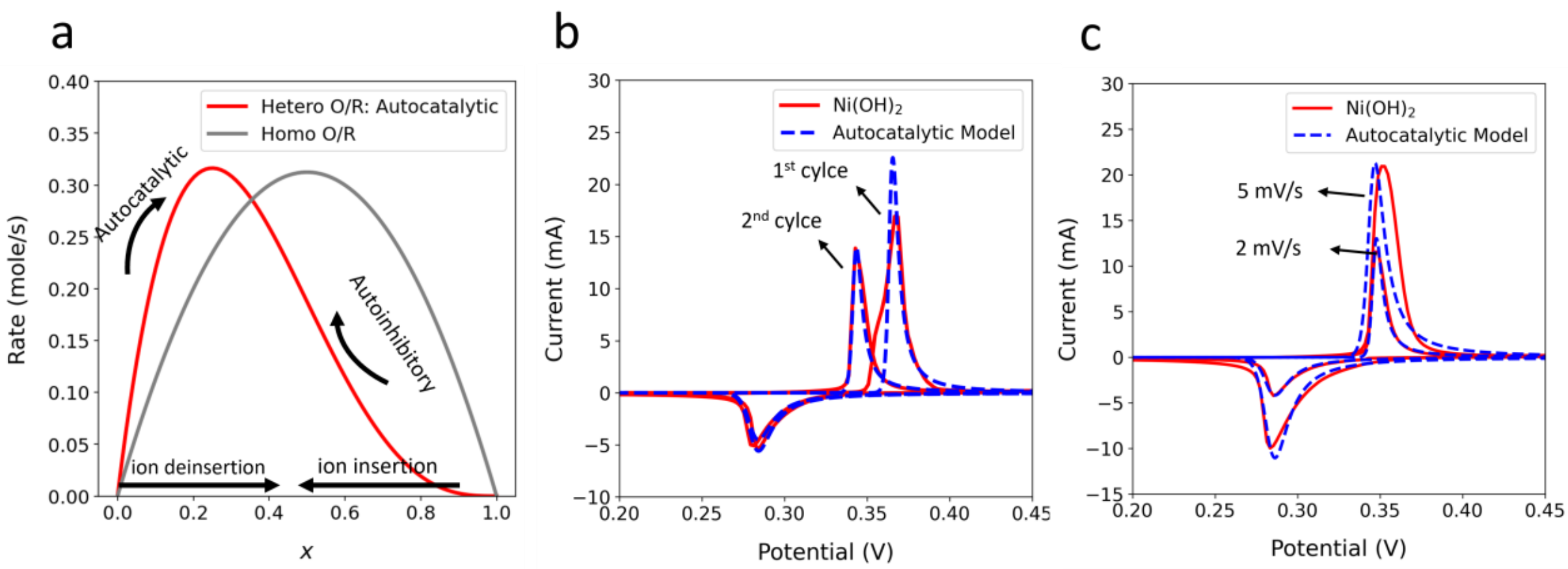

**Fig. 4. a)** Rate profiles of a homogenous (grey ) and heterogeneous (red) EPT based on a reaction-controlled model according to the rate equation $kx(1-x)^m$ with $m$ = 1 and 3, respectively, and $k_{insert} = k_{deinsert}$ for both systems. **b**) Comparison of experimental first and second CV cycles for $Ni(OH)_2$ thin film with the model with an initial mole fraction of O equal to $5\times10^{-8}$ and $2.6\times10^{-3}$**,** respectively, and **c)** Comparison of CV cycles at 2 and 5 mV/s with the model with $k_{deinsert}/k_{insert}$=2.5 (high reaction asymmetry) and $k_{deinsert}t/k_{insert}$ = 1.4 (low reaction asymmetry), respectively, in 0.1 KOH . Modified from ref. [29] with permission from the Royal Society of Chemistry.

Fig. 5 shows in situ X-ray microscopy images of $LiFePO_4$ at two different galvanostatic charge/discharge rates [33]. When the charge/discharge rate is low (0.15 C), the delithiation exhibits a clear phase separation as an intercalation wave, but the lithiation is more homogeneous across the particle surface area. Within this regime of charge-discharge rates, the process is reaction-controlled, featuring autocatalytic delithiation and autoinhibitory lithiation. However, when the charge-discharge rate is increased to 1.5C and 4C, the delithiation becomes more homogenous across the particle, hence reducing the reaction asymmetry. Thus, the phase transition behavior of $LiFePO_4$ is in agreement with that of $Ni(OH)_2$ presented in Fig.4.

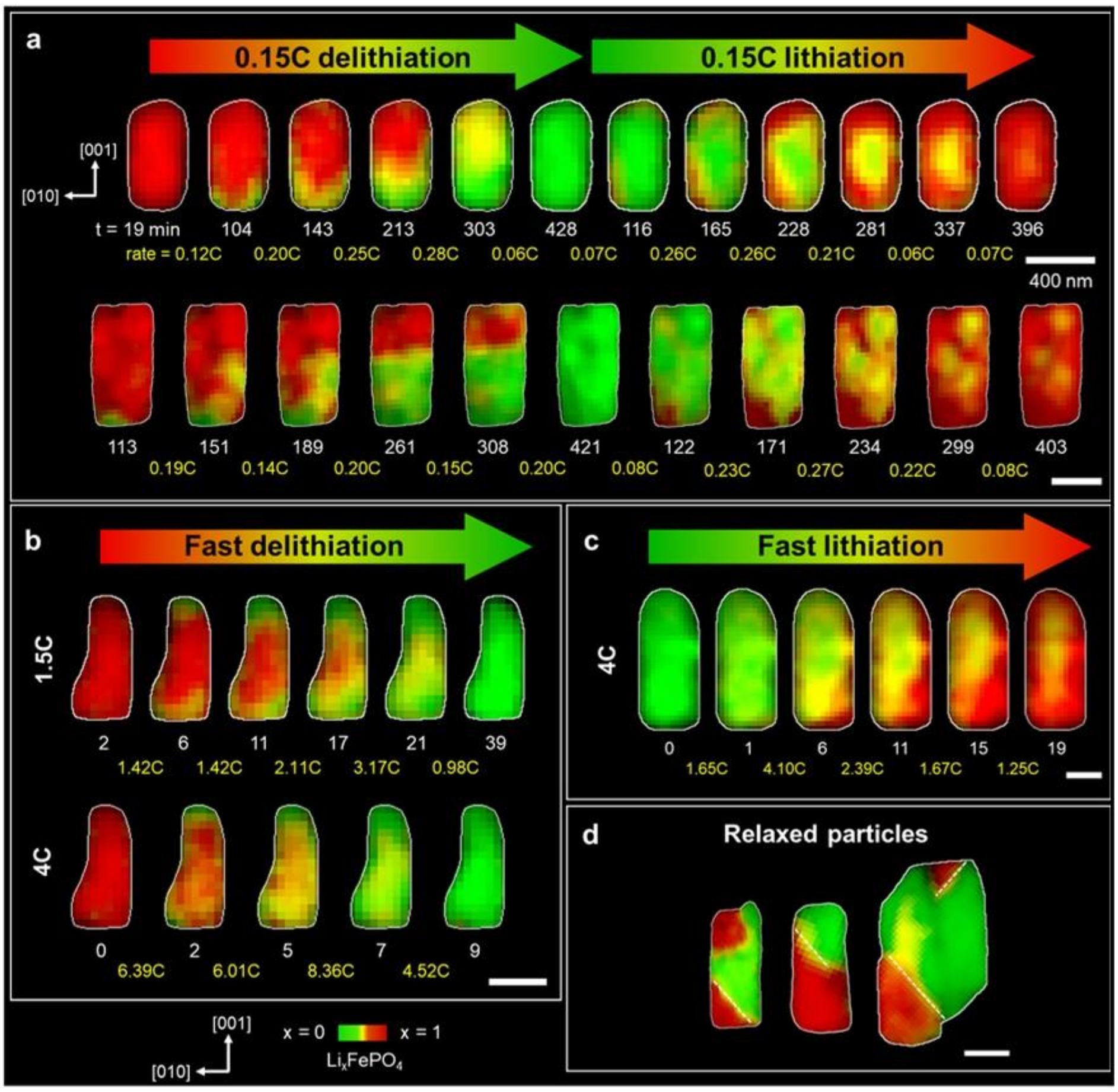

**Fig. 5** Operando X-ray scanning transmission microscopy of $LiFePO_4$ primary particles. Li composition maps of particles during (de)lithiation at rates of **(a)** 0.15C and **(b and c)** 1.5C and 4C. Red and green colors indicate Li-rich and Li-poor phases, respectively. White and yellow numbers are indicators of time **(d)** Ex situ Li composition maps of relaxed particles. The scale bar is 400 nm. Reproduced from ref. [33] With permission from the Royal Society of Chemistry

## Diffusion-limited EPT

When the EPT becomes influenced or limited by ion diffusion, at least two scenarios can be envisaged. 1) ion diffusion limitation within a homogeneous O/R film and 2) mixed effects of ion diffusion and phase transformation in the film. Here, we consider only the first case, which can principally be treated with a one-dimensional semi-infinite linear diffusion (SILD) along *y* axis as shown in scheme 1. If ion diffusion is limited within the redox film, which is usually the case for films with enough thickness, the EPT can be expressed based on eq. 3 and the following boundary conditions:

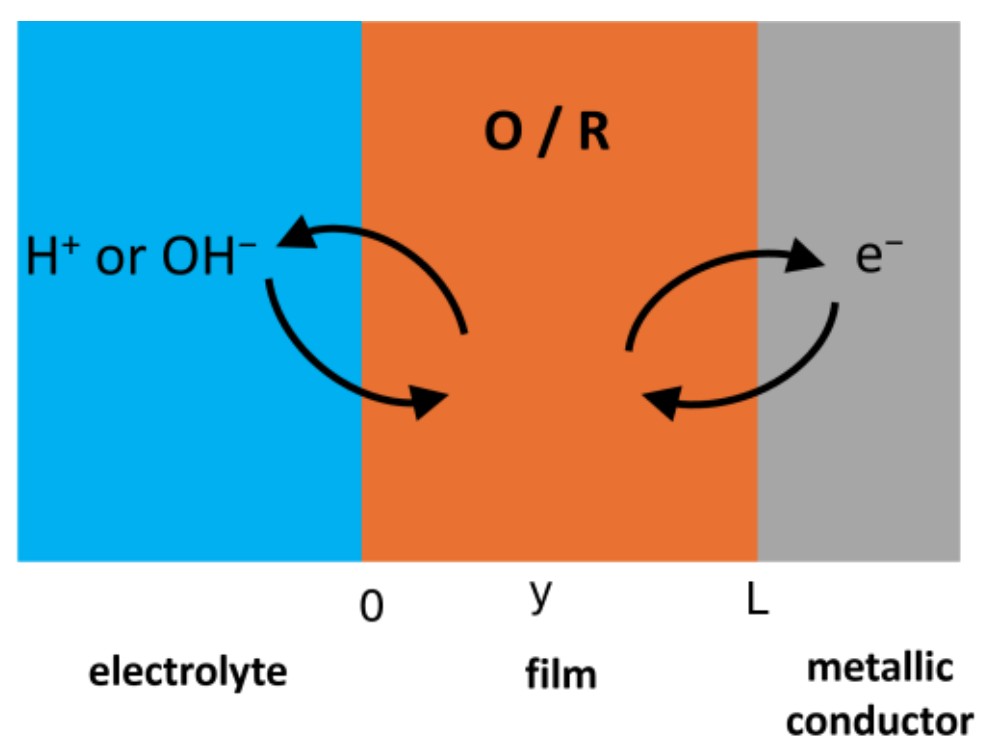

**Scheme 1.** A scheme of an ion-insertion electrode with thickness *L*, consisting of three phases: the aqueous electrolyte, the O/R film, and the metallic conductor.

$$\frac{\partial x(y,t)}{\partial t} = D\frac{\partial^2 x(y,t)}{\partial y^2} \qquad (3)$$

$$x(y,0) = 0 \qquad (4)$$

$$\lim_{y \to L} x(y,t) = 0 \qquad (5)$$

$$x(0,t) = \frac{\mathrm{e}^{0.5fE} - \frac{dx/dt}{k'}}{\mathrm{e}^{0.5fE} + \mathrm{e}^{-0.5fE}} \qquad (6)$$

Eq. 3 expresses Fick's second law based on the mole fraction of the O species (*x*). Eq. 4 and 5 result from the SILD assumption, and Eq. 6 simply results from rearranging the Butler-Volmer (B-V) equation, where $k'$ denotes the standard rate constant. For large values of $k'$, Eq. 6 reduces to the Nernst equation in eq. 1, as expected. The SILD condition can be realized for O/R films with high thickness and/or high charge-discharge rates [34]. The PDE in eq. 3 can be solved using the numerical solution of Nicholson and Schain (See the method section). Fig. 6. exhibits the resulting CV curves. Fig. 6a shows that the shape of the peaks and peak separation is jointly determined by the electron transfer standard rate constant and the diffusion coefficient within the film. For small enough values of $\sqrt{D}/k'$, the potentials of the CV curves follow the Nernst equation as exhibited in Fig. 6b and previously plotted in Fig. 2a (blue line) for different pH values.

It is interesting to compare these diffusion-limited CV curves with those of the diffusion-free CV curves in Fig. 3b. Here, the reduction and the oxidation tails at the potential limits do not tend to zero current (within experimental time) but rather a plateau due to the assumed SILD condition. Finally, these CV curves can be influenced by ohmic resistance, as well, witch is not considered here.

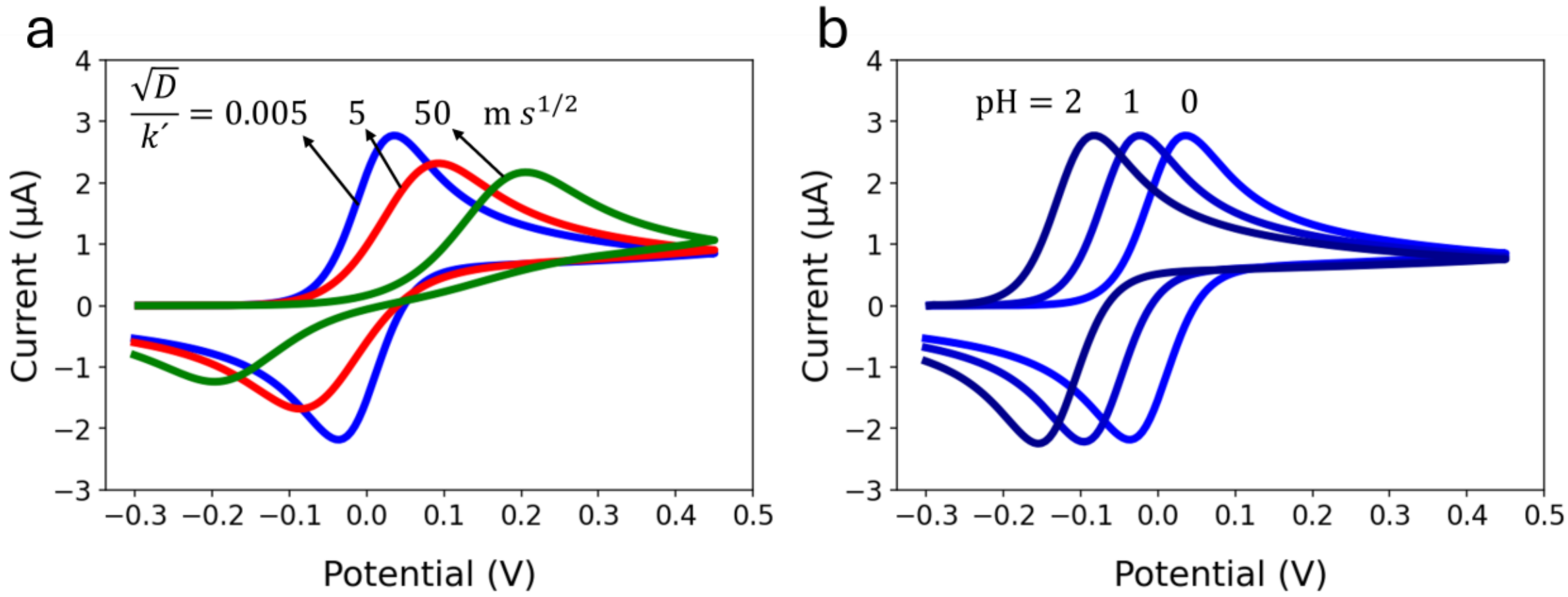

**Fig. 6** Simulation of CV curves based on the relative effects of the ion diffusion and the electron transfer kinetics. *D* and *k´* denote the diffusion coefficient and electron transfer rate constant in the O/R film.

An experimental example of an energy material that can be described by this framework is the ε-$MnO_2$ film [26,34]. The CV and the open-circuit potentials recorded for this electrode suggest a quasi-reversible proton/hydroxide ion insertion in aqueous solutions. For acidic solutions, the underlying EPT includes the $Mn^{2+}$/$MnO_2$ reaction, as well [26].

Finally, more comprehensive models have been recently reported for simulating the CV curves of lithium-intercalation electrodes via incorporating more fitting parameters, such as particle size and interaction parameter, into this diffusion framework (Fig. 6) [35–38]. Another factor that can further complicate the CV modelling of solid redox films arises when the redox film is not thick enough. Under these conditions, the CV model should incorporate the thickness and scan rate as well, such that it can simulate the transition from diffusion-limited framework to diffusion-free framework, as recently performed by Garay et al. [39,40].

## Conclusion and Future Prospects

Phase diagrams are basic tools for studying phase transitions and mixtures and have been widely applied to electrochemical phase transitions (EPTs). While straightforward to use, they fall short of capturing the complexities inherent in EPTs, particularly because they do not incorporate reaction dynamics or charge transfer in their conventional form. As a result, they offer limited insights into key features of EPTs such as kinetics, hysteresis, reaction asymmetry, and memory effects. Emerging research highlights further the influence of particle size[41], crystal structure, charge-discharge rates, and potential scan rate on reaction pathways—factors that phase diagrams do not address. Moreover, the effects of pH and electrolyte ions on EPT mechanisms, in particular on the CV current, remain largely unexplored. Implementing EPT models in non-aqueous batteries also requires accounting for the solid–electrode interphase (SEI), which significantly alters reaction mechanisms and kinetics. Consequently, there is currently no general framework to predict the electrochemical signals of EPTs based on structural parameters and experimental conditions.

Nucleation–growth models have provided valuable insights into the early stages of electrocrystallization, yet they face two main limitations in EPT studies. First, they are typically limited to initial phase formation and do not describe the growth of larger particles. Second, they generally treat only one direction of the transition—often the phase formation —while in real electrochemical systems, both directions, oxidation and reduction, are equally important. These constraints hinder their ability to explain anomalous behaviors observed in EPTs. Nonetheless, nucleation-growth models remain useful in EPT studies, as they are founded on fundamental thermodynamic principles. In addition, the application of empirical models like the Avrami equation is limited due to the lack of clear physical meaning behind their parameters and the underlying assumption that ion deinsertion/insertion (or metal dissolution/plating) phenomena are completely stochastic.

Finally, the current perspective attempted to reveal the more hidden sides of the EPTs, such as reaction hysteresis and asymmetry, using phase diagrams, electrochemical modeling, and platform EPT materials such as $Ni(OH)_2$ and $LiFePO_4$. We hope that it can motivate more fundamental research into EPTs. Extensive experimental data is needed to clarify the validity and scope of present models, and more advanced theories of EPT are needed to explain the dependence of EPTs on material structure and experimental parameters. An improved understanding of EPTs will pave the way for advancing current electrochemical energy technologies.

## Simulation Methods

All data analysis, visualization, and evaluations were carried out in Spyder, provided by Anaconda Inc., on a workstation with a CPU Intel(R) Core(TM) i5-13420H and 2.10 GHz and 16 GB of RAM. The simulation methods and the underlying model development for Figs 2, 3, and 4 are thoroughly described in refs. [25,29]. All potentials of experimental CVs are against an Ag/AgCl reference electrode with a negligible junction potential. All the experiments were carried out at 298 K, and this temperature was assumed in all the CV simulations. The values of the Faraday constant and the gas Constant are assumed to be 96485 C/mol and 8.314 J/(mol*K), respectively.

The simulation of the CV curves in Fig. 6 is performed as follows: The PDE in eq. 3 under the boundary conditions in eq. 4, 5, and 6 is solved with the numerical integration method of Nicholson and Shain [42]. In this method, the continuous integration variable $\tau$ is discretised as $\delta \times \mu$ denoting the length and number of steps, respectively. This results in a $g(n)$ function with $n = t/\delta$, that depends on its previous values as described by eq. 7:

$$g(n) = \frac{1}{1 + \frac{\sqrt{\pi D}}{4k'\sqrt{\delta}}\left(\mathrm{e}^{0.5fE} + \mathrm{e}^{-0.5fE}\right)}\left[\frac{\sqrt{\pi D/4\delta}}{1 + \mathrm{e}^{-fE}} + g(n-1) - g(1)\sqrt{n} - \mathrm{sum}\right] \quad (7)$$

Where the sum is equal to:

$$\mathrm{sum} = \sum_{i=1}^{n-2} \sqrt{n-i}\,[g(i+1) - g(i)] \quad (8)$$

$E$ is the potential program as defined by eq. 9 and 10 for forward and backward scans, respectively:

$$E = E_i - E_\mathrm{f} + vt \qquad 0 < t \leq \lambda \qquad (9)$$

$$E = E_i - E_\mathrm{f} + 2v\lambda - vt \qquad \lambda \leq t \qquad (10)$$

Where $v$ and $\lambda$ denote the scan rate and the reversal time, respectively.

The current can be expressed based on $g(t)$ as follows:

$$i(t) = \frac{\partial x(0,t)}{\partial t} q = g(n)q \qquad (11)$$

where $q$ is a scaling parameter corresponding to the charge (C) of the redox film being exchanged during a forward or backward scans.

Finally, if $\sqrt{D/k'}$ in Eq. 7 is very small , the system converges to the reversible EPT, which is simply described by the terms inside the brackets.

All the written Python programs used for the simulations in this work are freely available at https://github.com/malaiek/CV-SIM.git.

## Conflicts of Interest

There are no conflicts to declare.

## Acknowledgment

The author acknowledges the Alexander von Humboldt Foundation and the University of Greifswald.

## Table of Contents

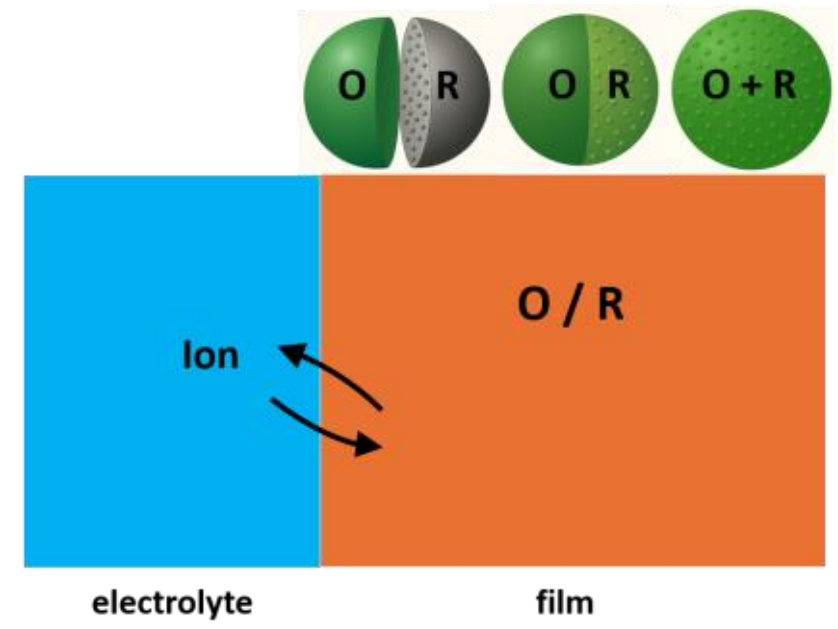

This Minireview explores the electrochemical thermodynamics, kinetics, and hysteresis of solid-to-solid electrochemical phase transitions (EPTs). It bridges phase transition thermodynamics of homogenous and heterogeneous systems with insertion electrochemistry of O/R films, highlighting the role of cyclic voltammetry in unraveling reaction mechanisms in energy materials.